\documentclass[10pt]{article}

\usepackage{graphicx}
\usepackage{amssymb}
\usepackage{amsmath}
\usepackage{hyperref}

\renewcommand{\Re}{\operatorname{Re}}
\renewcommand{\Im}{\operatorname{Im}}

\newcommand{\hc}[1]{#1^{\dagger}}

\newcommand{\abs}[1]{|#1|}

\newcommand{\tr}{\operatorname{tr}}
\newcommand{\diag}{\operatorname{diag}}
\newcommand{\adj}{\operatorname{adj}}

\title{Vacuum Stability Conditions and Potential Minima \\ for a Matrix Representation in Lightcone Orbit Space}
\author{Kristjan Kannike}
\date{}

\begin{document}
\maketitle

\begin{center}
\textit{National Institute of Chemical Physics and Biophysics, R\"{a}vala 10, Tallinn, Estonia}
\end{center}

\begin{abstract}
The orbit space for a scalar field in a complex square matrix representation obtains a Minkowski space structure from the Cauchy-Schwarz inequality. It can be used to find vacuum stability conditions and minima of the scalar potential. The method is suitable for fields such as a bidoublet, an $SU(2)$ triplet or $SU(3)$ octet. We use the formalism to find the vacuum stability conditions for the left-right symmetric potential of a bidoublet and left and right Higgs doublets.
\end{abstract}

\section{Introduction}
\label{sec:intro}

In extensions of the Standard Model (SM) scalar sector, it can be complicated to find vacuum stability conditions and to study the minimum structure of the scalar potential. For example, the analysis of the full scalar potential of the two-Higgs-doublet model (2HDM) \cite{Lee:1973iz} is rather involved. Fortunately, the potential depends only on a limited number of gauge invariants. The space of group invariants -- the orbit space -- generally has a non-trivial geometrical shape. But it is still simpler to analyse the orbit space than the space of all field components with its redundancies \cite{Michel:1971th,Abud:1981tf,Abud:1983id,Kim:1981xu,Kim:1983mc}. In particular, the orbit space of the 2HDM has a Minkowski space structure, because it resembles a forward lightcone in $1+3$ dimensions \cite{Ivanov:2006yq,Maniatis:2006fs}. The scalar quartic couplings form a Minkowski tensor and the mass terms gather in a four-vector. Consequently, tensor positivity conditions on the Minkowski space can be used to find vacuum stability conditions for the quartic couplings \cite{Ivanov:2006yq}. Furthermore, minima of the scalar potential can be analysed geometrically \cite{Degee:2009vp,Ivanov:2015nea}. The lightcone shape can be related to the Cauchy-Schwarz inequality \cite{Ivanov:2006yq}.

Another extension of the SM which benefits from an orbit space analysis is given by left-right symmetric models. The left-right gauge group $SU(2)_{L} \times SU(2)_{R} \times U(1)_{B-L}$ is a natural restoration of symmetry between left and right sectors \cite{Mohapatra:1974gc,Mohapatra:1974hk,Pati:1974yy,Senjanovic:1975rk}. The left-right symmetry can be spontaneously broken into the electroweak part of the SM gauge group either by Higgs doublets \cite{Senjanovic:1975rk,Mohapatra:1977be} or by  triplets \cite{Deshpande:1990ip,Maiezza:2016ybz} (which can also can explain neutrino mass through the seesaw mechanism \cite{Gell-Mann:1979vob,Glashow:1979nm,Minkowski:1977sc,Mohapatra:1979ia,Yanagida:1979as}). Previously, preliminary vacuum stability conditions for models with a bidoublet and triplets, but with most couplings set to zero, were given in \cite{BhupalDev:2018xya,Chakrabortty:2013mha}. Thereafter, a thorough study of vacuum stability for the left-right symmetry broken by a bidoublet and triplets was made in \cite{Chauhan:2019fji}. Recent work with left-right doublets includes \cite{Gabrielli:2016vbb} and \cite{Babu:2020bgz}.

We observe that the orbit space of a scalar field in a complex square matrix representation, if the quartic scalar potential can be written in terms of two invariants, also looks like a $1+2$ dimensional forward lightcone. The Minkowski structure arises from the Cauchy-Schwarz inequality for the inner product of matrices. We then use positivity of the quartic coupling tensor to derive necessary and sufficient vacuum stability or bounded-from-below conditions for the self-couplings of the matrix field by analogy with the 2HDM. Portal couplings to the Higgs boson can be can be presented as a Minkowski vector \cite{Alanne:2016wtx}. If the couplings are real, we can reduce the vacuum stability problem to copositivity \cite{Kannike:2012pe}, i.e. positivity on positive vectors. The technique can be applied to various scalar fields, such as an $SU(2)_{L}$ triplet, or an $SU(3)$ sextet or octet.

We apply the formalism to derive the vacuum stability conditions on a left-right-symmetric scalar potential with a bidoublet and left and right Higgs doublets \cite{Senjanovic:1975rk,Mohapatra:1977be}. The conditions for the bidoublet self-couplings are straightforward to obtain. They are equivalent to those previously published in another form in \cite{Chauhan:2019fji}. For real bidoublet self-couplings and Higgs portal couplings, the problem reduces to copositivity, and we obtain necessary or sufficient vacuum stability conditions for the full potential.

The lightcone orbit space for a matrix is described and vacuum stability conditions derived in Section~\ref{sec:lightcone}. Portal couplings with the Higgs doublet are added in Section~\ref{sec:higgs:coupling}. The scalar potential is minimised in Section~\ref{sec:minim}. In Section~\ref{sec:left:right} we derive analytical vacuum stability conditions for a left-right symmetric model with a bidoublet and left-right Higgs doublets. We conclude in Section \ref{sec:concl}.

\section{Lightcone Orbit Space from the Cauchy-Schwarz Inequality}
\label{sec:lightcone}

\subsection{Matrix Self-Coupling Potential and the Lightcone Orbit Space}
\label{sec:self:lightcone}

For two matrices $A$ and $B$ of suitable dimensions, the inner product is defined as $\tr (\hc{A} B)$.
Given a scalar field in a complex square matrix representation $M$ of a (gauge) group, the  invariants $\tr \hc{M} M$ and $\tr M^{2}$ satisfy the Cauchy-Schwarz inequality
\begin{equation}
  \tr M^{\dagger 2} \tr M^{2} \leqslant (\tr \hc{M} M)^{2}.
\label{eq:matrix:Cauchy-Schwarz}
\end{equation}
We assume that these are the only independent group invariants needed to write the scalar potential.
Let us express these field bilinears in terms of real variables $r^{\mu}$ with $\mu = 0, 1, 2$:
\begin{equation}
  \tr \hc{M} M = r^{0}, \quad \tr M^{2} = r^{1} + i r^{2}.
\end{equation}
We can now write the Cauchy-Schwarz inequality \eqref{eq:matrix:Cauchy-Schwarz} as 
\begin{equation}
  (r^{0})^{2} - (r^{1})^{2} - (r^{2})^{2} \geqslant 0,
\label{eq:matrix:lightcone}
\end{equation}
which, together with $r^{0} \geqslant 0$, describes the orbit space of the scalar field $M$ as a forward lightcone $\text{LC}^{+}$ in $1+2$ dimensions.\footnote{Another possible parametrisation is $\tr \hc{M} M = r$, $\tr M^{2} = r \rho e^{i \phi}$ with $r \geqslant 0$, $0 \leqslant \rho \leqslant 1$, $0 \leqslant \phi < 2 \pi$. To derive vacuum stability conditions for the self-couplings, one can then demand that the minimum value of the coefficient of $r^{4}$ in the quartic part of the potential be positive. The resulting conditions, however, are somewhat less concise.}  An $SO(1,2)$ Lorentz transformation will leave the inequality~\eqref{eq:matrix:lightcone} intact. In particular, an $SO(2)$ rotation of the $(r^{1}, r^{2})$ `spatial' vector by the angle $\theta$ corresponds to the $U(1)$ phase rotation $M \to e^{i \theta/2} M$, and a boost in the direction of $r^{1}$ with rapidity $\varphi$ to $M \to M \cosh (\varphi/2) + \hc{M} \sinh (\varphi/2)$. For convenience, we will use relativistic terminology with obvious meanings of `time-like', `space-like' etc.

The mass terms and quartic self-couplings of the matrix field are given by
\begin{equation}
\begin{split}
  V_{M} &= \mu^{2}_{M} \tr \hc{M} M + \frac{1}{2} \left( \mu^{\prime 2}_{M} \tr M^{2} + \mu^{\prime 2 *}_{M} \tr M^{\dagger 2} \right)
  + \lambda_{M} (\tr \hc{M} M)^{2} + \lambda'_{M} \tr M^{\dagger 2} \tr M^{2}
  \\
  &+ \frac{1}{2} \left( \lambda_{M}^{\prime\prime} \left(\tr M^{2} \right)^{2} 
  + \lambda_{M}^{\prime\prime *} \left(\tr M^{\dagger 2} \right)^{2} \right)
  + \frac{1}{2} \tr \hc{M} M \left( \lambda_{M}^{\prime\prime\prime} \tr M^{2} 
  + \lambda_{M}^{\prime\prime\prime *} \tr M^{\dagger 2} \right),
\end{split}
\label{eq:potential:M}
\end{equation}
where the parameters $\mu^{\prime 2}_{M}$, $\lambda''_{M}$ and $\lambda'''_{M}$ can be complex.

In terms of the lightcone variables $r^{\mu}$, the potential \eqref{eq:potential:M} can be written as
\begin{equation}
  V_{M} = \mu^{2}_{M\mu} r^{\mu} + r^{\mu} \lambda_{\mu\nu} r^{\nu},
\label{eq:potential:1+2}
\end{equation}
where the mass vector
\begin{equation}
  \mu^{2}_{M\mu} = \left( \mu^{2}_{M}, \Re \mu^{\prime 2}_{M}, -\Im \mu^{\prime 2}_{M}  \right)
\end{equation}
and the quartic coupling tensor
\begin{equation}
  \lambda_{\mu\nu} = 
  \begin{pmatrix}
    \lambda_{M} & \frac{1}{2} \Re{\lambda'''_{M}} & -\frac{1}{2} \Im\lambda'''_{M}
    \\
   \frac{1}{2} \Re \lambda'''_{M} & \lambda'_{M} + \Re \lambda''_{M} & -\Im \lambda''_{M}
    \\
    -\frac{1}{2} \Im \lambda'''_{M} & -\Im \lambda''_{M} & \lambda'_{M} - \Re \lambda''_{M}
  \end{pmatrix}.
\label{eq:coupling:tensor:1+1}
\end{equation}
The quartic coupling tensor \eqref{eq:coupling:tensor:1+1} can be diagonalised by an $SO(1,2)$ Lorentz transformation since all such transformations are available from the fundamental theory. (In some models, such as a three-Higgs-doublet model (3HDM), for example, $\lambda_{\mu\nu}$ is not always diagonalisable, because the 3HDM orbit space does not fill the whole forward lightcone and hence not all Lorentz transformations are available \cite{Ivanov:2010ww}.) The diagonalised tensor has the form $\lambda_{\mu\nu}^{\mathrm{D}} = \diag (\Lambda_{0}, -\Lambda_{1}, -\Lambda_{2})$, where the minus signs of the space-like eigenvalues arise from the pseudo-Euclidean metric.

\subsection{Vacuum Stability Conditions for the Matrix Self-Couplings}
\label{sec:self:vacuum}

In order for the matrix self-coupling potential $V_{M}$ to be bounded from below, the potential must be positive in the limit of large fields, in which we can ignore terms with mass dimensions and take into account only the quartic part of the potential.\footnote{We consider strict positivity of the quartic potential. In the case of non-strict positivity, the quartic potential may have flat directions, in which the mass and cubic potential must be positive if the potential is to be bounded from below.} Therefore, the quartic coupling tensor $\lambda_{\mu\nu}$ has to be positive on the forward lightcone \eqref{eq:matrix:lightcone}. For this, its eigenvalues have to satisfy \cite{Ivanov:2006yq}
\begin{equation}
  \Lambda_{0} > 0, \quad \Lambda_{0} > \Lambda_{1}, \quad \Lambda_{0} > \Lambda_{2}.
\label{eq:Lorentz:tensor:pos}
\end{equation}
We will analyse in detail only the most interesting case of real couplings: then the coupling tensor is 
\begin{equation}
  \lambda_{\mu\nu} = 
  \begin{pmatrix}
    \lambda_{M} & \frac{1}{2} \lambda'''_{M} & 0
    \\
   \frac{1}{2} \lambda'''_{M} & \lambda'_{M} + \lambda''_{M} & 0
    \\
    0 & 0 & \lambda'_{M} - \lambda''_{M}
  \end{pmatrix}.
\label{eq:coupling:tensor:1+1:real}
\end{equation}
More generally, the tensor with complex couplings \eqref{eq:coupling:tensor:1+1} can also be brought into a similar block-diagonal form by a phase rotation of the field $M$, choosing $\arg \lambda'''_{M} = (1/2) \arg \lambda''_{M}$ without loss of generality and then doing a rotation in the $r^{1}r^{2}$-plane by the angle $-\arg \lambda'''_{M}$. Such a choice of coupling phases, however, is somewhat unusual.

For $\lambda'''_{M} = 0$, the eigenvalues of the coupling tensor \eqref{eq:coupling:tensor:1+1:real}  are directly given by $\Lambda_{0} = \lambda_{00} = \lambda_{M}$, $-\Lambda_{1} = \lambda_{11} = \lambda'_{M} + \lambda''_{M}$ and $-\Lambda_{2} = \lambda_{22} = \lambda'_{M} - \lambda''_{M}$, so the conditions \eqref{eq:Lorentz:tensor:pos} yield
\begin{equation}
  \lambda_{M} > 0, \quad \lambda_{M} + \lambda'_{M} + \lambda''_{M} > 0, \quad \lambda_{M} + \lambda'_{M} - \lambda''_{M} > 0.
  \label{eq:pos:cond:diag:1+2}
\end{equation}
These conditions are quite intuitive, considering that the minimum of the potential is achieved for a negative $-\Lambda_{1}$ or $-\Lambda_{2}$, if the field lies on the surface of the lightcone. Depending on the values of these couplings, either of the last two conditions in Eq. \eqref{eq:pos:cond:diag:1+2} may dominate. In particular, since for real couplings, $r^{2}$ appears in only the $\lambda_{22} (r^{2})^{2}$ term, we see that for $-\Lambda_{2} \geqslant 0$, we must choose $r^{2} = 0$ to minimise the potential. On the other hand, if $-\Lambda_{2} < 0$, it is most convenient to consider it as a function of $r^{0}$ and $r^{1}$ and the potential is minimised when it takes its value on the lightcone, i.e. $(r^{2})^{2} = (r^{0})^{2} - (r^{1})^{2}$.

In case of $\lambda'''_{M} \neq 0$, the tensor \eqref{eq:coupling:tensor:1+1:real} can be diagonalised by the Lorentz transformation
\begin{equation}
  \lambda = 
  \begin{pmatrix}
    \cosh \varphi & \sinh \varphi & 0
    \\
    \sinh \varphi & \cosh \varphi & 0
    \\
    0 & 0 & 1
  \end{pmatrix}.
\end{equation}
There are four solutions to the equation $\Lambda^{\rho}_{\;\; \mu} \Lambda^{\sigma}_{\;\; \nu} \lambda_{\rho\sigma} = \lambda_{\mu\nu}^{\mathrm{D}}$ for the rapidity $\varphi$, but three of them are spurious, since they do not give an identity Lorentz transformation for $\lambda'''_{M} = 0$, that is, for a coupling tensor that already is diagonal. The physical solution, 
\begin{equation}
  \varphi = \frac{1}{4} \ln \frac{\lambda_{M} + \lambda'_{M} + \lambda''_{M} - \lambda'''_{M}}{\lambda_{M} + \lambda'_{M} + \lambda''_{M} + \lambda'''_{M}},
\end{equation}
yields
\begin{align}
  \Lambda_{0} &= \frac{1}{2} \left( \lambda_{M} - \lambda'_{M}  - \lambda''_{M}
  + \sqrt{(\lambda_{M} + \lambda'_{M} + \lambda''_{M} - \lambda'''_{M})(\lambda_{M} + \lambda'_{M} + \lambda''_{M} + \lambda'''_{M})} \right),
  \\
  -\Lambda_{1} &= \frac{1}{2} \left( -\lambda_{M} + \lambda'_{M} + \lambda''_{M}
  + \sqrt{(\lambda_{M} + \lambda'_{M} + \lambda''_{M} - \lambda'''_{M})(\lambda_{M} + \lambda'_{M} + \lambda''_{M} + \lambda'''_{M})} \right),
  \\
  -\Lambda_{2} &= \lambda'_{M} - \lambda''_{M}.
\end{align}

The positivity conditions \eqref{eq:Lorentz:tensor:pos} for the coupling tensor \eqref{eq:coupling:tensor:1+1:real} are then given, after simplification, by
\begin{align}
  \lambda_{M} + \lambda'_{M} + \lambda''_{M} - \abs{\lambda'''_{M}}&> 0,
\label{eq:pos:cond:gen:1+1:1}
  \\
  \lambda_{M} - \lambda'_{M} - \lambda''_{M} 
  + \sqrt{(\lambda_{M} + \lambda'_{M} + \lambda''_{M} - \lambda'''_{M})(\lambda_{M} + \lambda'_{M} + \lambda''_{M} + \lambda'''_{M})} &> 0,
\label{eq:pos:cond:gen:1+1:2}
  \\
  \lambda_{M} + \lambda'_{M} - 3 \lambda''_{M} 
  + \sqrt{(\lambda_{M} + \lambda'_{M} + \lambda''_{M} - \lambda'''_{M})(\lambda_{M} + \lambda'_{M} + \lambda''_{M} + \lambda'''_{M})} &> 0.
\label{eq:pos:cond:gen:1+1:3}
\end{align}
For $\lambda'''_{M} = 0$, these conditions reduce to Eq. \eqref{eq:pos:cond:diag:1+2} as required; for $\lambda'''_{M} \neq 0$, the conditions \eqref{eq:pos:cond:diag:1+2} are necessary. Note that for $-\Lambda_{2} = \lambda_{22} = \lambda'_{M} - \lambda''_{M} > 0$, the second condition \eqref{eq:pos:cond:gen:1+1:2} is stronger than the last one \eqref{eq:pos:cond:gen:1+1:3}, since a positive $-\Lambda_{2}$ only takes us away from the minimum; for $-\Lambda_{2} < 0$, it is the opposite. Thus we can subsume the condition $\Lambda_{0} > 0$ \eqref{eq:pos:cond:gen:1+1:2} into $\Lambda_{0} - \Lambda_{2} > 0$ \eqref{eq:pos:cond:gen:1+1:3} by making the substitution $-\Lambda_{2} \to -\Lambda_{2} \, \theta(\Lambda_{2})$ in the latter, where $\theta$ is the Heavyside step function. In summary, the self-coupling potential \eqref{eq:potential:1+2} is bounded from below if the conditions \eqref{eq:pos:cond:gen:1+1:1}, \eqref{eq:pos:cond:gen:1+1:2} and \eqref{eq:pos:cond:gen:1+1:3} are satisfied.\footnote{Notice that the conditions \eqref{eq:pos:cond:gen:1+1:1}, \eqref{eq:pos:cond:gen:1+1:2} and \eqref{eq:pos:cond:gen:1+1:3} are similar to the vacuum stability conditions of the self-couplings of a complex singlet \cite{Kannike:2012pe}. This is not an accident, since for a complex singlet $S = (s_{R} + i s_{I})/\sqrt{2}$, we can also write its quartic potential, if its self-couplings are real, in terms of lightcone variables $r^{0} = (s_{R}^{2} + s_{I}^{2})/2$ and $r^{1} = (s_{R}^{2} - s_{I}^{2})/2$, which satisfy $r^{0} \geqslant 0$ and $(r^{0})^{2} - (r^{1})^{2} \geqslant 0$.}

\begin{figure}[tb]
\begin{center}
  \includegraphics{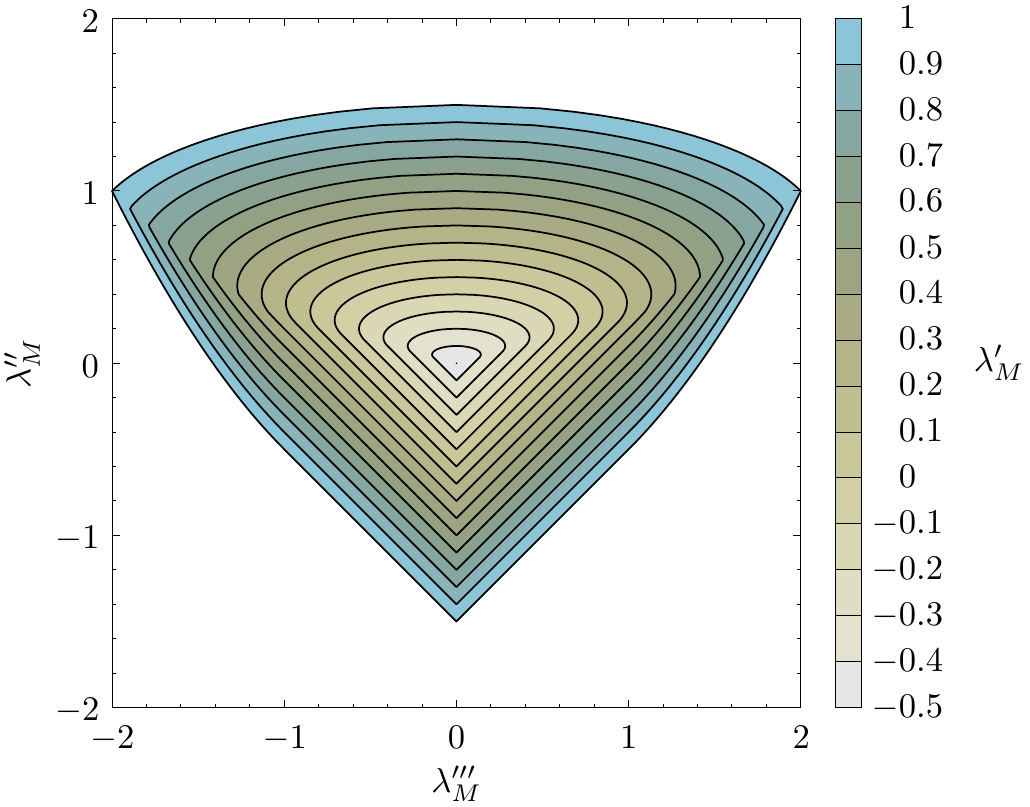}
\caption{Allowed parameter space for the matrix self-couplings from vacuum stability for $\lambda_{M} = 0.5$ and $-1 \leqslant \lambda'_{M} \leqslant 1$. The tip of the `mountain' is at $\lambda'''_{M} = \lambda''_{M} = 0$, $\lambda'_{M} = -0.5$.}
\label{eq:matrix:self}
\end{center}
\end{figure}

The vacuum stability conditions for the bidoublet self-couplings are demonstrated in Figure~\ref{eq:matrix:self} in the $\lambda''_{M}$ vs. $\lambda'''_{M}$ plane for $\lambda_{M} = 0.5$ and $-1 \leqslant \lambda'_{M} \leqslant 1$. The allowed region, whose shape resembles an inverted mountain, is colour-coded for $\lambda'_{M}$. The tip of the `mountain' is at $\lambda'''_{M} = \lambda''_{M} = 0$, $\lambda'_{M} = -0.5$.

\section{Couplings to the Higgs Doublet}
\label{sec:higgs:coupling}

\subsection{Higgs Portal Couplings and the Full Scalar Potential}
\label{sec:higgs:portal}

In a realistic model, we couple the matrix field $M$ to the Higgs doublet $H$.\footnote{Of course, in a dark sector, a similar potential can describe the interactions of the scalar $M$ and a Standard Model singlet $S$ with $\abs{S}^{2}$ substituted for $\abs{H}^{2}$.} First of all, the Higgs mass term and self-coupling are given by 
\begin{equation}
  V_{H} = \mu^{2}_{H} \abs{H}^{2} + \lambda_{H} \abs{H}^{4}.
\end{equation}
We can write couplings of the field $M$ to the Higgs doublet as
\begin{equation}
  V_{HM} = \lambda_{HM} \abs{H}^{2} \tr \hc{M} M + \frac{1}{2} \abs{H}^{2}  \left( \lambda'_{HM} \tr M^{2} + \lambda_{HM}^{\prime *} \tr M^{\dagger 2} \right),
\end{equation}
which we can write in the form
\begin{equation}
  V_{HM} = \lambda_{H\mu} r^{\mu} \abs{H}^{2},
\end{equation}
where
\begin{equation}
   \lambda_{H\mu} = (\lambda_{HM}, \Re \lambda'_{HM}, -\Im \lambda'_{HM}).
\end{equation}

The full scalar potential is given by
\begin{equation}
  V = V_{H} + V_{HM} + V_{M} = \mu^{2}_{H} \abs{H}^{2} + \mu^{2}_{M\mu} r^{\mu} + \lambda_{H} \abs{H}^{4} + \lambda_{H\mu} r^{\mu} \abs{H}^{2}  + r^{\mu} \lambda_{\mu\nu} r^{\nu}.
\label{eq:V}
\end{equation}
In general, there could be other terms in the potential, e.g. as a cubic term given by the determinant of $M$, for example. Such terms are a fly in the ointment: they do not fit straight away in our parametrisation, although they can be written via the lightcone variables at the expense of introducing additional orbit space parameters.

\subsection{Vacuum Stability Conditions for the Full Scalar Potential}
\label{sec:vacuum:full}

For the full scalar potential \eqref{eq:V} of $H$ and $M$, the vacuum stability conditions \eqref{eq:pos:cond:gen:1+1:1}, \eqref{eq:pos:cond:gen:1+1:2} and \eqref{eq:pos:cond:gen:1+1:3} for the $M$ self-coupling potential \eqref{eq:potential:1+2} are necessary. Likewise, one has to require $\lambda_{H} > 0$. In order to find the full necessary and sufficient conditions for vacuum stability, the Higgs portal couplings $\lambda_{H\mu}$ must be taken into account. If $\lambda_{H\mu}$ is in the forward lightcone, that is, $\lambda_{HM} \geqslant 0$, $\lambda_{HM} \geqslant \abs{\lambda'_{HM}}$, then $\lambda_{H\mu} r^{\mu}$ is positive in the whole forward lightcone and nothing need be done. On the other hand, if $\lambda_{H\mu}$ is in the backward lightcone, then $\lambda_{H\mu} r^{\mu}$ is negative in the whole forward lightcone. In this case, we can minimise the quartic part of the potential \eqref{eq:V} over $\abs{H}^{2}$,
\begin{equation}
  \abs{H}^{2} = -\frac{\lambda_{H\mu} r^{\mu}}{2 \lambda_{H}},
\end{equation}
essentially substituting
\begin{equation}
  \lambda_{\mu\nu} \to \lambda_{\mu\nu} - \frac{\lambda_{H\mu} \lambda_{H\nu}}{4 \lambda_{H}}
\end{equation}
in the conditions \eqref{eq:Lorentz:tensor:pos}. If $\lambda_{H\mu}$ is space-like, however, then $\lambda_{H\mu} r^{\mu}$ is negative only in a part of the forward lightcone and demanding positivity over the whole forward lightcone only yields a sufficient, not necessary condition for the vacuum stability of the potential.
 
There is a considerable simplification if we restrict ourselves to the case of real couplings, i.e. no explicit CP-violation. In this case, instead of trying to look for a complicated condition for a space-like $\lambda_{H\mu}$, we will sidestep this issue altogether. We will  reduce the problem of vacuum stability to copositivity \cite{Kannike:2012pe}, that is, positivity on positive vectors.

For real couplings, the $\lambda_{22} (r^{2})^{2}$ term remains as the only potential term with $r^{2}$. As before, if $\lambda_{22} = \lambda'_{M} - \lambda''_{M} \geqslant 0$, the $\lambda_{22}$ term will give a non-negative contribution to the potential and can be ignored in finding vacuum stability conditions. On the other hand, if $\lambda_{22} < 0$, then the potential is minimised when it takes the value on the lightcone, i.e. $(r^{2})^{2} = (r^{0})^{2} - (r^{1})^{2}$. This means that we must require
\begin{align}
  \lambda'_{M} - \lambda''_{M} \geqslant 0 & \implies \left. V \right|_{(r^{2})^{2} = 0} > 0,
  \label{eq:vacuum:la22:pos}
  \\
  \lambda'_{M} - \lambda''_{M} < 0 & \implies \left. V \right|_{(r^{2})^{2} = (r^{0})^{2} - (r^{1})^{2}} > 0,
   \label{eq:vacuum:la22:neg}
\end{align}
in the limit of large field values; the implication $p \implies q$ is equivalent to $\lnot p \lor q$.
Imposing $r^{2} = 0$ or the lightcone condition $(r^{2})^{2} = (r^{0})^{2} - (r^{1})^{2}$ means, in effect, that the coupling tensor $\lambda_{\mu\nu}$ is reduced to its upper-left block with $\mu,\nu = 0,1$ and, in the latter case, $\lambda_{00} \to \lambda_{00} + \lambda_{22}$ and $\lambda_{11} \to \lambda_{11} - \lambda_{22}$. As a shorthand for the two implications \eqref{eq:vacuum:la22:pos} and \eqref{eq:vacuum:la22:neg}, we can multiply the $\lambda_{22}$ coupling by the Heaviside step function $\theta(-\lambda_{22})$. Having minimised over $r^{2}$, we are left with a potential that depends only on $r^{0}$, $r^{1}$ and $\abs{H}^{2}$. While $r^{0}$ and $r^{1}$ are physical on the $1+1$ forward lightcone $\mathrm{LC}^{+}$, the square of the Higgs doublet $\abs{H}^{2}$ is physical on non-negative numbers $\mathbb{R}_{+}$, so the whole orbit space is $\mathrm{LC}^{+} \! \times \mathbb{R}_{+}$. In order to take into account all the quartic couplings, we augment the reduced $\lambda_{\mu\nu}$ with couplings to the Higgs boson and the Higgs self-coupling. The resulting tensor, in the basis $(r^{0}, r^{1}, \abs{H}^{2})$, is given by
\begin{equation}
  \lambda_{AB} = 
  \begin{pmatrix}
    \lambda_{M} + (\lambda'_{M} - \lambda''_{M}) \, \theta(\lambda''_{M} - \lambda'_{M}) 
    & \frac{1}{2} \lambda'''_{M} 
    & \frac{1}{2} \lambda_{HM} 
    \\
    \frac{1}{2} \lambda'''_{M} & \lambda'_{M} +\lambda''_{M} 
    - (\lambda'_{M} - \lambda''_{M}) \, \theta(\lambda''_{M} - \lambda'_{M})
    & \frac{1}{2} \lambda'_{HM}
    \\
    \frac{1}{2} \lambda_{HM} 
    & \frac{1}{2} \lambda'_{HM}
    & \lambda_{H}
  \end{pmatrix},
\end{equation}
where the mixed indices $A, B = \mu, H$ with $\mu = 0,1$.

\begin{figure}[tb]
\begin{center}
\includegraphics{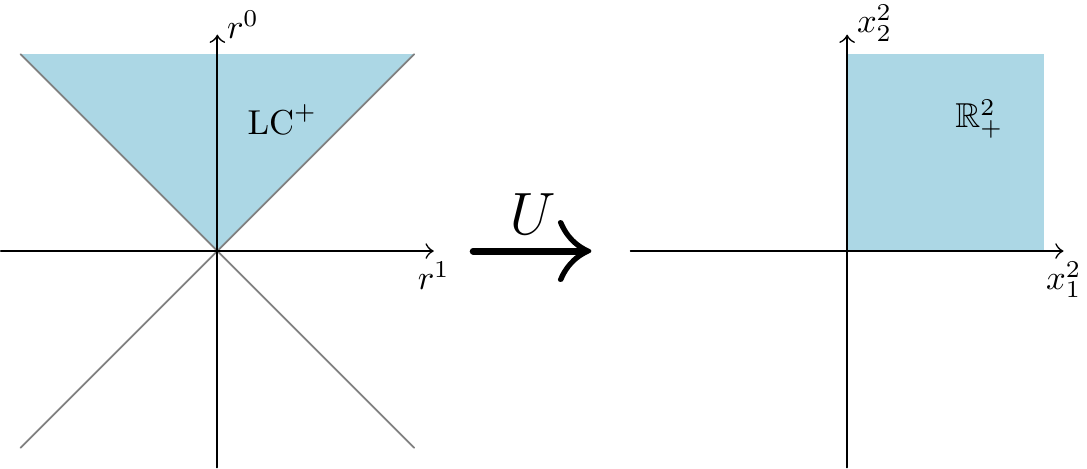}
\caption{Rotation of the $1+1$ forward lightcone $\mathrm{LC}^{+}$ into the non-negative quadrant $\mathbb{R}^{2}_{+}$.}
\label{fig:lc:pq}
\end{center}
\end{figure}

By rotating the  $1+1$ forward lightcone $\text{LC}^{+}$ into the non-negative quadrant $\mathbb{R}^{2}_{+}$ of the $r^{0}r^{1}$-plane, as illustrated in Figure~\ref{fig:lc:pq}, the whole orbit space will transform from the mixed $\mathrm{LC}^{+} \! \times \mathbb{R}_{+}$ into the non-negative octant $\mathbb{R}_{+}^{3}$. Positivity on the lightcone is then reduced to copositivity. The Higgs boson and the portal couplings are taken into account in the same fashion.  In particular, we rotate the tensor $\lambda_{AB}$ by
\begin{equation}
  U^{A}_{\;\;\: i} = 
  \begin{pmatrix}
  \frac{1}{\sqrt{2}} & \frac{1}{\sqrt{2}} & 0
  \\
  -\frac{1}{\sqrt{2}} & \frac{1}{\sqrt{2}} & 0
  \\
  0 & 0 & 1
  \end{pmatrix},
\end{equation}
where the lowercase Latin indices denote usual Cartesian coordinates. The $\lambda_{AB}$ tensor becomes, upon the rotation, the quartic coupling matrix given by
\begin{equation}
  \lambda_{ij} = (U^{T})^{A}_{\;\;\: i} \lambda_{AB} U^{B}_{\;\;\: j}.
\end{equation}

The full potential, minimised over $r^{2}$, can thus be written as
\begin{equation}
  V = x^{2}_{i} \lambda_{ij} x^{2}_{j}
\end{equation}
in the basis
\begin{equation}
 x_{1}^{2} = \frac{r^{0} - r^{1}}{\sqrt{2}}, \quad x_{2}^{2} = \frac{r^{0} + r^{1}}{\sqrt{2}}, \quad x_{3}^{2} = \abs{H}^{2}.
\label{eq:basis:xi:sq}
\end{equation}
The physical orbit space is then given by non-negative $x_{i}^{2}$, that is, $\mathbb{R}_{+}^{3}$.
The coupling matrix is given by
\begin{equation}
  \lambda_{ij} = 
  \begin{pmatrix}
    \lambda_{-} &
    \frac{1}{2} \lambda_{\mp}
    & \frac{1}{2} \lambda_{H-}
    \\
   \frac{1}{2} \lambda_{\mp}
    & \lambda_{+} & \frac{1}{2} \lambda_{H+}
    \\
    \frac{1}{2} \lambda_{H-} & \frac{1}{2} \lambda_{H+} & \lambda_{H}
  \end{pmatrix},
\label{eq:coupling:real:matrix:H}
\end{equation}
where
\begin{align}
  \lambda_{-} &= \frac{1}{2} (\lambda_{M} + \lambda'_{M} + \lambda''_{M} - \lambda'''_{M}),
  \\
  \lambda_{+} &= \frac{1}{2} (\lambda_{M} + \lambda'_{M} + \lambda''_{M} + \lambda'''_{M}),
  \\
  \lambda_{\mp} &=  \lambda_{M} - \lambda'_{M} - \lambda''_{M} + 2 (\lambda'_{M} - \lambda''_{M}) \, \theta(\lambda''_{M} - \lambda'_{M}),
  \\
  \lambda_{H-} &= \frac{1}{\sqrt{2}} (\lambda_{HM} - \lambda'_{HM}),
  \\
  \lambda_{H+} &= \frac{1}{\sqrt{2}} (\lambda_{HM} + \lambda'_{HM}).
\end{align}

The necessary and sufficient vacuum stability conditions for the potential \eqref{eq:V} with real couplings are obtained by requiring copositivity of the matrix \eqref{eq:coupling:real:matrix:H} and are given by \cite{Kannike:2012pe}
\begin{equation}
\begin{split}
  \lambda_{H} > 0, \qquad\qquad \lambda_{-} > 0, \qquad\qquad \lambda_{+} &> 0,
  \\
  \bar{\lambda}_{H-} = \frac{1}{2} \lambda_{H-} + \sqrt{ \lambda_{H} \lambda_{-} } > 0,
  \qquad
  \bar{\lambda}_{H+} = \frac{1}{2} \lambda_{H+} +\sqrt{ \lambda_{H} \lambda_{+} } > 0,
  \qquad
  \bar{\lambda}_{\mp} = \frac{1}{2} \lambda_{\mp} + \sqrt{ \lambda_{-} \lambda_{+} } &> 0,
  \\
  \sqrt{\lambda_{H} \lambda_{-} \lambda_{+}} + \lambda_{H-} \sqrt{\lambda_{+}} + \lambda_{H+} \sqrt{\lambda_{-}} + \lambda_{\mp} \sqrt{\lambda_{H}} 
  + \sqrt{2 \bar{\lambda}_{H-} \bar{\lambda}_{H+} \bar{\lambda}_{\mp}} &> 0.
  \end{split}
  \label{eq:gen:cop}
\end{equation}
The vacuum stability conditions \eqref{eq:pos:cond:gen:1+1:1}, \eqref{eq:pos:cond:gen:1+1:2} and \eqref{eq:pos:cond:gen:1+1:3} for the self-couplings of $M$ are reproduced by the conditions $\lambda_{-} > 0$, $\lambda_{+} > 0$ and $\bar{\lambda}_{\mp} > 0$.

\section{Potential Minimisation}
\label{sec:minim}

The stationary points of the scalar potential \eqref{eq:V} are given by solving 
\begin{align}
  0 &= \frac{\partial V}{\partial \abs{H}^{2}} \frac{\partial \abs{H}^{2}}{\partial H}, 
  \label{eq:grad:H:dagger}
  \\ 
  0 &= \frac{\partial V}{\partial r^{\mu}} \frac{\partial r^{\mu}}{\partial M}.
  \label{eq:grad:r:mu}
\end{align}
Multiplying Eq.~\eqref{eq:grad:H:dagger} from the right by $H$, we obtain
\begin{equation}
  0 = \abs{H}^{2} (\mu^{2}_{H} + \lambda_{H\mu} r^{\mu} + 2 \lambda_{H} \abs{H}^{2}),
  \label{eq:grad:H:dagger:scalar}
\end{equation}
which is solved by $\abs{H}^{2} = 0$ (the trivial solution) or 
\begin{equation}
  \abs{H}^{2} = -\frac{1}{2 \lambda_{H}} (\mu^{2}_{H} + \lambda_{H\mu} r^{\mu}).
\end{equation}

Similarly to Eq. \eqref{eq:grad:H:dagger:scalar}, we also want to take Eq.~\eqref{eq:grad:r:mu} into a form that does not explicitly depend on the matrix $M$, but only on the lightcone variables $r^{\mu}$. To that end, we multiply Eq.~\eqref{eq:grad:r:mu} by $M^{T}$, take a trace and the real part:
\begin{equation}
\begin{split}
  0 &= \frac{1}{2} \tr \left[ \frac{\partial V}{\partial r^{\mu}} \frac{\partial r^{\mu}}{\partial M} M^{T}
  + \left( \frac{\partial V}{\partial r^{\mu}} \frac{\partial r^{\mu}}{\partial M} M^{T} \right)^{*} \right]
  \\
  &=  \frac{\partial V}{\partial r^{\mu}} \Re \left( \tr  \frac{\partial r^{\mu}}{\partial M} M^{T} \right)
  \\
  &= \frac{\partial V}{\partial r^{\mu}} r^{\mu},
\end{split}
\label{eq:min:M:1}
\end{equation}
where we used Eqs. \eqref{eq:d:r:d:M:M:1}, \eqref{eq:d:r:d:M:M:2} and \eqref{eq:d:r:d:M:M:3} from Appendix \ref{sec:deriv} in the last step. Similarly, by multiplying Eq.~\eqref{eq:grad:r:mu} by $M^{*}$, taking the imaginary part, or both, we additionally obtain
\begin{align}
  0 &= \frac{\partial V}{\partial r^{1}} r^{2} - \frac{\partial V}{\partial r^{2}} r^{1},
  \label{eq:min:M:2}
  \\
  0 &= \frac{\partial V}{\partial r^{0}} r^{1} + \frac{\partial V}{\partial r^{1}} r^{0},
  \label{eq:min:M:3}
  \\
  0 &= \frac{\partial V}{\partial r^{0}} r^{2} + \frac{\partial V}{\partial r^{2}} r^{0}.
  \label{eq:min:M:4}
\end{align}

The tip of the lightcone $r^{\mu} = 0$ trivially satisfies all the equations.
For $r^{2} = 0$, Eqs. \eqref{eq:min:M:2} and \eqref{eq:min:M:4} are identically zero. 
For real couplings, which is the case we study, we can eliminate $r^{2}$ as before by choosing either $r^{2} = 0$ (for $\lambda_{22} \geqslant 0$) or $(r^{2})^{2} = (r^{0})^{2} - (r^{1})^{2}$ (for $\lambda_{22} < 0$) in the minimum (note that this may not hold in extrema other than minima). This eliminates Eqs. \eqref{eq:min:M:2} and \eqref{eq:min:M:4}, and Eq.~\eqref{eq:min:M:1} becomes
\begin{equation}
  0= \frac{\partial V}{\partial r^{0}} r^{0} + \frac{\partial V}{\partial r^{1}} r^{1}.
  \label{eq:min:M:1:alt}
\end{equation}

All Eqs. \eqref{eq:min:M:1}, \eqref{eq:min:M:2}, \eqref{eq:min:M:3} and \eqref{eq:min:M:4} are satisfied if the partial derivatives of $V$ with respect to $r^{\mu}$ vanish (whether all couplings are real or not):
\begin{equation}
  0 = \frac{\partial V}{\partial r^{\mu}} = \mu^{2}_{M\mu} + \lambda_{H\mu} \abs{H}^{2} + 2 \lambda_{\mu\nu} r^{\nu}.
  \label{eq:r:mu:eq}
\end{equation}
Such solutions, if $\abs{H}^{2} = 0$, are given by
\begin{equation}
  r^{\mu} = -\frac{1}{2} (\lambda^{-1})^{\mu \nu} \mu^{2}_{M \nu},
  \label{eq:r:mu:sol}
\end{equation}
which exists if the coupling tensor $\lambda_{\mu\nu}$ is non-singular, i.e. its eigenvalues are not zero. If one or more eigenvalues vanish, then the solution \eqref{eq:r:mu:sol} is given by
\begin{equation}
  r^{\mu} = -\frac{1}{2} (\tilde{\lambda}^{-1})^{\mu \nu} \tilde{\mu}^{2}_{M \nu} + \rho^{\mu},
  \label{eq:r:mu:sol:restr}
\end{equation}
where $(\tilde{\lambda}^{-1})^{\mu \nu}$ and $\tilde{\mu}^{2}_{M \nu}$ are restrictions to the orthogonal subspace with non-vanishing eigenvalues and $\rho^{\mu}$ belongs to the kernel of $\lambda_{\mu\nu}$, i.e. $\lambda_{\mu\nu} \rho^{\nu} = 0$.
With non-zero $\abs{H}^{2}$, the solutions have the same structure, with the substitutions
\begin{equation}
  \mu^{2}_{M\mu} \to \mu^{2}_{M\mu} - \frac{\mu^{2}_{H}}{2 \lambda_{H}} \lambda_{H\mu}, 
  \qquad
  \lambda_{\mu\nu} \to \lambda_{\mu\nu} - \frac{1}{4} \lambda_{H\mu} \lambda_{H\nu}
\end{equation}
in Eqs. \eqref{eq:r:mu:sol} and \eqref{eq:r:mu:sol:restr}. 

Another type of solution arises if the partial derivatives are non-zero. One sees immediately from Eqs. \eqref{eq:min:M:3} and \eqref{eq:min:M:1:alt} that then $r^{1} = \pm r^{0}$, so both solutions are on the lightcone (but in general have different magnitude).\footnote{Note that in the $x_{i}^{2}$ basis used in Section \ref{sec:vacuum:full}, these solutions correspond to $x_{1}^{2} = 0$ or $x_{2}^{2} = 0$.} The solutions are obtained from
\begin{equation}
  \frac{\partial V}{\partial r^{0}} = \pm\frac{\partial V}{\partial r^{1}}.
  \label{eq:pm:sols}
\end{equation}
Explicit minimum solutions are presented in Appendix~\ref{sec:minima}.

The mass matrix can be calculated in the basis $(M, M^{\dagger}, H, \hc{H})$, where $M$ is shorthand for $M_{ij}$, that is, $ij$ can be treated as a multi-index; similarly, $H$ stands for $H_{i}$. E.g. the second derivative $[\partial^{2}/(\partial M \partial \hc{M})]_{ij,kl} = \partial^{2}/(\partial M_{ij} \partial \hc{M}_{kl})$ and $[\partial M/\partial M]_{ij,kl} = \partial M_{ij}/\partial M_{kl}$. Then, for example, we have, as a diagonal block of the mass matrix,
\begin{equation}
\begin{split}
  m^{2}_{\hc{M}M} = \frac{\partial^{2}V}{\partial \hc{M} \partial M} 
  &= \frac{\partial}{\partial \hc{M}} \left( \frac{\partial V}{\partial r^{\mu}} 
  \frac{\partial r^{\mu}}{\partial M} \right)
  \\
  &=  \frac{\partial^{2} V}{\partial r^{\mu} \partial r^{\nu}} \frac{\partial r^{\mu}}{\partial \hc{M}} \frac{\partial r^{\nu}}{\partial M} + \frac{\partial V}{\partial r^{\mu}}  \frac{\partial^{2} r^{\mu}}{\partial \hc{M}\partial M},
\end{split}
\end{equation}
where $\partial r^{\mu}/\partial M$ are given by \eqref{eq:d:r:d:M}.

The eigenvalues of the mass matrix can be given in terms of the lightcone variables $r^{\mu}$ and the square of the Higgs doublet $\abs{H}^{2}$. The invariants of the mass matrix, such as its trace, determinant, and so on that enter the characteristic polynomial can be calculated in matrix notation. In practice it may be easier, however, to first calculate the eigenvalues in terms of the elements of $M$ and $\hc{M}$ and only then express them via $r^{\mu}$.

\section{Left-Right Model with a Bidoublet and Higgs Doublets}
\label{sec:left:right}

Left-right symmetry is an extension of the SM gauge group that restores the parity symmetry at high energies. Before spontaneous symmetry breaking, the left- and right-handed fermions are treated in the same way. The left-right gauge group is $SU(3)_{C} \times SU(2)_{L} \times SU(2)_{R} \times U(1)_{B-L}$. We consider vacuum stability and minimum structure for the model with left and right Higgs doublets and a left-right bidoublet \cite{Senjanovic:1975rk,Mohapatra:1977be}. The scalar fields and their irreducible gauge representations of the model are  given in Table~\ref{tab:field}. The fields transform as
\begin{equation}
  \Phi \to U_{L} \Phi \hc{U_{R}}, \quad H_{L} \to U_{L} H_{L}, \quad H_{R} \to U_{R} H_{R}
\end{equation}
under $SU(2)_{L} \times SU(2)_{R}$ for the gauge transformations $U_{L} \in SU(2)_{L}, U_{R} \in SU(2)_{R}$. One can also define $\tilde{\Phi} = \sigma_{2} \Phi^{*} \sigma_{2}$, which transforms in the same way as $\Phi$.

The electric charge has the form
\begin{equation}
  Q = T_{3L} + T_{3R} + \frac{B-L}{2},
\end{equation}
where $T_{3}$ is the third component of the weak isospin.

\begin{table}[tb]
\caption{Scalar fields and their representations in the left-right model.}
\begin{center}
\begin{tabular}{c|cccc}
  Fields & $SU(3)_{C}$ & $SU(2)_{L}$ & $SU(2)_{R}$ & $U(1)_{B-L}$
  \\
  \hline
  $\Phi = \begin{pmatrix} \phi_{1}^{0} & \phi_{1}^{+} \\ \phi_{2}^{-} & \phi_{2}^{0}
  \end{pmatrix}$ & $\mathbf{1}$ & $\mathbf{2}$ & $\mathbf{2}$ & $0$ 
  \\
  $H_{L} = \begin{pmatrix} H_{L}^{+} \\ H_{L}^{0} \end{pmatrix}$ & $\mathbf{1}$ & $\mathbf{2}$ & $\mathbf{1}$ & $1$
  \\
  $H_{R} = \begin{pmatrix} H_{R}^{+} \\ H_{R}^{0} \end{pmatrix}$ & $\mathbf{1}$ & $\mathbf{1}$ & $\mathbf{2}$ & $1$
\end{tabular}
\end{center}
\label{tab:field}
\end{table}%

The scalar potential can be written as 
\begin{equation}
  V = V_{\Phi} + V_{H} + V_{H\Phi},
\label{eq:pot}
\end{equation}
where the bidoublet potential, comprising mass terms and quartic self-couplings, is given by
\begin{equation}
\begin{split}
V_{\Phi} &= \mu_{1}^{2} \tr \hc{\Phi} \Phi + \mu_{2}^{2} (\tr \tilde{\Phi} \hc{\Phi} 
  +  \tr \hc{\tilde{\Phi}} \Phi)
+ \lambda_{1} (\tr \hc{\Phi} \Phi)^{2} 
  + \lambda_{2} \left[ (\tr \tilde{\Phi} \hc{\Phi})^{2} + (\tr \hc{\tilde{\Phi}} \Phi)^{2} \right]
    \\
  &+ \lambda_{3} (\tr \tilde{\Phi} \hc{\Phi}) (\tr \hc{\tilde{\Phi}} \Phi) 
  + \lambda_{4} (\tr \hc{\Phi} \Phi) \left[ \tr \tilde{\Phi} \hc{\Phi} + \tr \hc{\tilde{\Phi}} \Phi \right],
\end{split}
\label{eq:bidoublet}
\end{equation}
the mass terms and quartic scalar interactions among the Higgs doublets are given by
\begin{equation}
  V_{H} = \mu_{L}^{2} \abs{H_{L}}^{2} + \mu_{R}^{2} \abs{H_{R}}^{2} + \lambda_{L} \abs{H_{L}}^{4} + \lambda_{R} \abs{H_{R}}^{4} 
  + \lambda_{LR} \abs{H_{L}}^{2} \abs{H_{R}}^{2},
\end{equation}
and the interactions between the bidoublet and the doublets are given by
\begin{equation}
\begin{split}
  V_{H\Phi} &= \mu (\hc{H_{L}} \Phi H_{R} + \hc{H_{R}} \hc{\Phi} H_{L}) 
  + \tilde{\mu} (\hc{H_{L}} \tilde{\Phi} H_{R} + \hc{H_{R}} \hc{\tilde{\Phi}} H_{L})
  + \lambda_{\Phi L} \tr \hc{\Phi} \Phi \abs{H_{L}}^{2}
  \\
  & 
  + \tilde{\lambda}_{\Phi L} (\tr \tilde{\Phi} \hc{\Phi} + \tr \hc{\tilde{\Phi}} \Phi) \abs{H_{L}}^{2} + \lambda_{\Phi R} \tr \hc{\Phi} \Phi \abs{H_{R}}^{2}
  + \tilde{\lambda}_{\Phi R} (\tr \tilde{\Phi} \hc{\Phi} + \tr \hc{\tilde{\Phi}} \Phi) \abs{H_{R}}^{2}
 \\
 &+ \lambda'_{\Phi L} \hc{H_{L}} \Phi \hc{\Phi} H_{L} 
 + \lambda'_{\Phi R} \hc{H_{R}} \hc{\Phi} \Phi H_{R} 
 + \tilde{\lambda}'_{\Phi L} \hc{H_{L}} \tilde{\Phi} \hc{\tilde{\Phi}} H_{L} 
 + \tilde{\lambda}'_{\Phi R} \hc{H_{R}} \hc{\tilde{\Phi}} \tilde{\Phi} H_{R}.
\end{split}
\label{eq:V:HS}
\end{equation}
While several interaction couplings, such as $\lambda_{4}$, for example, could be complex, we take them real, so that there is no explicit CP-violation.

\subsection{Bidoublet Lightcone and Vacuum Stability}
\label{sec:left:right:bidoublet}

We express the bidoublet gauge invariant bilinears, as in Section \ref{sec:self:lightcone}, via lightcone variables:
\begin{equation}
  \tr \hc{\Phi} \Phi = r^{0}, \quad \tr \hc{\tilde{\Phi}} \Phi = r^{1} + i r^{2}.
\end{equation}
Notice that while the bidoublet is a general complex $2 \times 2$ matrix, it is sufficient to consider a diagonal $\Phi$ to reproduce any point in the lightcone.

In terms of the lightcone variables, the bidoublet potential \eqref{eq:bidoublet} is
\begin{equation}
  V_{\Phi} = \mu^{2}_{\Phi\mu} r^{\mu} + r^{\mu} \lambda_{\mu\nu} r^{\nu},
  \label{eq:bidoublet:self:coupling:pot:mib}
\end{equation}
where the mass term vector is
\begin{equation}
  \mu^{2}_{\Phi\mu} = (\mu^{2}_{1}, 2 \mu^{2}_{2}, 0)
\label{eq:bidoublet:mass}
\end{equation}
and the quartic coupling tensor is
\begin{equation}
  \lambda_{\mu\nu} = 
  \begin{pmatrix}
    \lambda_{1} & \lambda_{4} & 0 \\
    \lambda_{4} & \lambda_{3} + 2 \lambda_{2} & 0     \\
   0 & 0     & \lambda_{3} - 2 \lambda_{2}
  \end{pmatrix}.
\label{eq:couplings:bid:lor}
\end{equation}

For $\lambda_{4} = 0$, the eigenvalues of the coupling matrix \eqref{eq:couplings:bid:lor} in the minimal integrity basis are directly given by $\Lambda_{0} = \lambda_{1}$, $-\Lambda_{1} = \lambda_{3} + 2 \lambda_{2}$ and $-\Lambda_{2} = \lambda_{3} - 2 \lambda_{2}$. The conditions \eqref{eq:Lorentz:tensor:pos} for the positivity of a Lorentz tensor give
\begin{equation}
  \lambda_{1} > 0, \quad \lambda_{1} + \lambda_{3} - 2 \abs{\lambda_{2}} > 0.
  \label{eq:cop:bid}
\end{equation}

By comparing the bidoublet quartic self-coupling tensor \eqref{eq:couplings:bid:lor} with the generic self-coupling tensor \eqref{eq:coupling:tensor:1+1:real}, we obtain -- from Eqs. \eqref{eq:pos:cond:gen:1+1:1}, \eqref{eq:pos:cond:gen:1+1:2} and \eqref{eq:pos:cond:gen:1+1:3} -- that the vacuum stability conditions for the bidoublet self-couplings are
\begin{align}
  \lambda_{1} + 2 \lambda_{2} + \lambda_{3} - 2 \abs{\lambda_{4}} &> 0,
\label{eq:pos:cond:bid:1+1:1}
  \\
  \lambda_{1} - 2 \lambda_{2} - \lambda_{3} + 
  \sqrt{(\lambda_{1} + 2 \lambda_{2} + \lambda_{3})^2 - 4 \lambda_{4}^2} &> 0,
\label{eq:pos:cond:bid:1+1:2}
  \\
  \lambda_{1} - 6 \lambda_{2} + \lambda_{3} + 
  \sqrt{(\lambda_{1} + 2 \lambda_{2} + \lambda_{3})^2 - 4 \lambda_{4}^2} &> 0.
\label{eq:pos:cond:bid:1+1:3}
\end{align}
It can be shown that Eqs. \eqref{eq:pos:cond:bid:1+1:1}, \eqref{eq:pos:cond:bid:1+1:1} and \eqref{eq:pos:cond:bid:1+1:3} are equivalent to the vacuum stability conditions for the bidoublet self-couplings previously given in Ref. \cite{Chauhan:2019fji} in another form.

\subsection{Full Left-Right Scalar Potential in Lightcone Variables}
\label{sec:left:right:bidoublet:doublet}

\begin{figure}[tb]
\begin{center}
\includegraphics{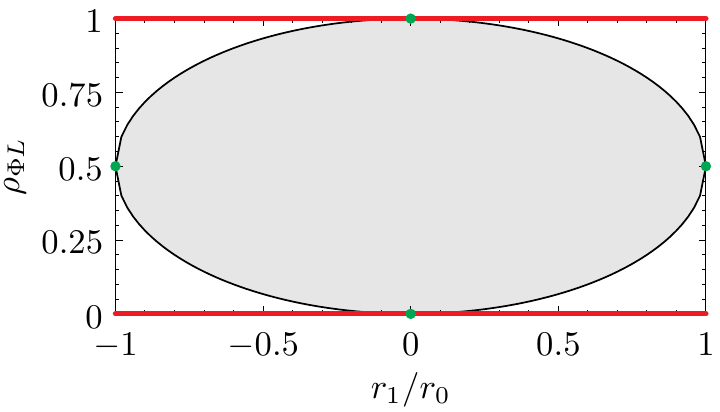}
\caption{Orbit space relation between $\rho_{\Phi L}$ and $r_{1}/r_{0}$. A similar parameter space is available for $\rho_{\Phi R}$ vs. $r_{1}/r_{0}$. Considering green points yields simplest necessary conditions and considering red lines sufficient conditions for vacuum stability.}
\label{fig:rho:Phi:L:r1:r0}
\end{center}
\end{figure}

Altogether, the potential \eqref{eq:pot} is then
\begin{equation}
\begin{split}
  V &= \mu_{L}^{2} \abs{H_{L}}^{2} + \mu_{R}^{2} \abs{H_{R}}^{2} + \mu^{2}_{\Phi\mu} r^{\mu} + r^{\mu} \lambda_{\mu\nu} r^{\nu} + \lambda_{L\mu} \abs{H_{L}}^{2} r^{\mu}
  + \lambda_{R\mu} \abs{H_{R}}^{2} r^{\mu} 
  \\
  &+ \lambda_{L} \abs{H_{L}}^{4} + \lambda_{R} \abs{H_{R}}^{4} 
  + \lambda_{LR} \abs{H_{L}}^{2} \abs{H_{R}}^{2}
  \\
  & + \sqrt{\abs{H_{L}}^{2} \abs{H_{R}}^{2} r^{0}} (\mu \rho \cos \theta + \tilde{\mu} \tilde{\rho} \cos \tilde{\theta}),
\end{split}
\label{eq:bidoublet:full:potential}
\end{equation}
where the vector $\mu^{2}_{\Phi\mu}$ is given by Eq. \eqref{eq:bidoublet:mass}, the tensor $\lambda_{\mu\nu}$ is given by Eq.~\eqref{eq:couplings:bid:lor} and
\begin{align}
  \lambda_{L \mu} &= (\lambda_{\Phi L} + \rho_{\Phi L} \lambda'_{\Phi  L} + \tilde{\rho}_{\Phi L} \tilde{\lambda}'_{\Phi  L}, 2 \tilde{\lambda}_{\Phi L}, 0),
  \\
  \lambda_{R \mu} &= (\lambda_{\Phi R} + \rho_{\Phi R} \lambda'_{\Phi R} + \tilde{\rho}_{\Phi R} \tilde{\lambda}'_{\Phi  R}, 2 \tilde{\lambda}_{\Phi R}, 0).
\end{align}
We have used $\hc{H_{L}} \Phi H_{R} = \sqrt{\abs{H_{L}}^{2} \abs{H_{R}}^{2} r^{0}} \rho e^{i \theta}$ with $0 \leqslant \rho \leqslant 1$ and $0 \leqslant \theta < 2 \pi$ and a similar parametrisation for the second trilinear term with $\tilde{\rho}$ and $\tilde{\theta}$ in the cubic terms. For the quartic terms, we have used $\hc{H_{L}} \Phi \hc{\Phi} H_{L} = \rho_{\Phi L} r^{0} \abs{H_{L}}^{2}$ and $\hc{H_{R}} \hc{\Phi} \Phi H_{R} = \rho_{\Phi R} r^{0} \abs{H_{R}}^{2}$ with $0 \leqslant \rho_{\phi L}, \rho_{\phi R} \leqslant 1$ for the $\lambda'_{\Phi L}$ and $\lambda'_{\Phi R}$ terms, and a similar parametrisation with $\tilde{\rho}_{\Phi L}$ and $\tilde{\rho}_{\Phi R}$ for the $\tilde{\lambda}'_{\Phi L}$ and $\tilde{\lambda}'_{\Phi R}$ terms.
In fact, it is easy to see, considering a basis where the bidoublet $\Phi$ is diagonal, that
\begin{equation}
  \rho_{\Phi L} + \tilde{\rho}_{\Phi L} = 1, \qquad \rho_{\Phi R} + \tilde{\rho}_{\Phi R} = 1,
  \label{eq:rho:rel}
\end{equation}
while $\rho_{\Phi L}$ and $\rho_{\Phi R}$ are independent of each other.
In addition, physical values for $\rho_{\Phi L,R}$ are within the ellipse given by
\begin{equation}
  \left(\frac{r_{1}}{r_{0}} \right)^{2} + (2 \rho_{\Phi L,R} - 1)^{2} = 1
 \label{eq:orbit:ellipse}
\end{equation}
illustrated in Figure~\ref{fig:rho:Phi:L:r1:r0}. 

\subsection{Vacuum Stability Conditions}
\label{sec:left:right:vacuum}

In order to derive vacuum stability conditions for the potential \eqref{eq:bidoublet:full:potential}, we follow the procedure of Section~\ref{sec:vacuum:full}. First of all, we minimise the potential with respect to $r^{2}$, in effect substituting $\lambda_{00} = \lambda_{1} \to \lambda_{1} + (\lambda_{3} - 2 \lambda_{2}) \theta(2 \lambda_{2} - \lambda_{3})$ and $\lambda_{11} = 2 \lambda_{2} + \lambda_{3} \to 2 \lambda_{2} + \lambda_{3} - (\lambda_{3} - 2 \lambda_{2}) \theta(2 \lambda_{2} - \lambda_{3})$. 

Due to the non-trivial dependency on $\rho_{\Phi L}$ and $\rho_{\Phi R}$ in Eq.~\eqref{eq:orbit:ellipse}, derivation of the full necessary and sufficient vacuum stability conditions becomes very complicated. It is straightforward, however, to write down some necessary \emph{or} sufficient conditions. Because the potential depends on $\rho_{\Phi L,R}$ linearly, it is minimised when these parameters take extremal values on the boundary of the ellipse \eqref{eq:orbit:ellipse}.

First of all, if we set $r^{1}$ to zero, then $\rho_{\Phi L,R}$ can vary in their whole ranges. Since $r^{0} \geqslant 0$, we can immediately use copositivity constraints in the $(r^{0}, \abs{H_{L}}^{2}, \abs{H_{R}}^{2})$ basis. Similarly, if we set $r^{1} = \pm r^{0}$, we can set $\rho_{\Phi L} = \rho_{\Phi R} = 1/2$ and do the same. Together, these choices correspond to the green points in the ends of the semiaxes of the ellipse in Figure~\ref{fig:rho:Phi:L:r1:r0}. In fact, we can set $r^{1} = \pm k r^{0}$ to get a necessary condition for $\rho_{\Phi L}$ and $\rho_{\Phi R}$ on the boundaries of the ellipse for a constant $k$. The above two choices correspond to $k = 0$ and $k = 1$, respectively. The coupling matrix in the $(r^{0}, \abs{H_{L}}^{2}, \abs{H_{R}}^{2})$ basis is given by
\begin{equation}
  \lambda_{k} = 
  \begin{pmatrix}
    \lambda_{00} & \frac{1}{2} \lambda_{L0} & \frac{1}{2} \lambda_{R0}
  \\
    \frac{1}{2} \lambda_{L0} & \lambda_{L} & \frac{1}{2} \lambda_{LR}
  \\
    \frac{1}{2} \lambda_{R0} & \frac{1}{2} \lambda_{LR} & \lambda_{R}
  \end{pmatrix},
\end{equation}
where
\begin{align}
  \lambda_{00} &= \lambda_{1} + k^{2} (2 \lambda_{2} + \lambda_{3}) + (1- k^{2}) ( \lambda_{3} - 2 \lambda_{2}) \theta(2 \lambda_{2} - \lambda_{3}) \pm 2 k \lambda_{4},
  \\
  \lambda_{L0} &= \lambda_{\Phi L} + \lambda'_{\Phi L} \rho_{\Phi L} + \tilde{\lambda}'_{\Phi L} (1- \rho_{\Phi L}) \pm 2 k \tilde{\lambda}_{\Phi L},
  \\
  \lambda_{R0} &=  \lambda_{\Phi R} + \lambda'_{\Phi R} \rho_{\Phi R} + \tilde{\lambda}'_{\Phi R} (1- \rho_{\Phi R}) \pm 2 k \tilde{\lambda}_{\Phi R},
\end{align}
where we have taken into account the relations \eqref{eq:rho:rel} to substitute for $\tilde{\rho}_{\Phi L}$ and $\tilde{\rho}_{\Phi R}$. We must separately consider the four combinations of signs in the solution to Eq.~\eqref{eq:orbit:ellipse},
\begin{equation}
  \rho_{\Phi L} = \frac{1}{2} (1 \pm \sqrt{1 - k^{2}}), \quad   \rho_{\Phi R} = \frac{1}{2} (1 \pm \sqrt{1 - k^{2}}),
\end{equation}
in addition to the two signs of $\pm k$.

The resulting necessary conditions for the left-right symmetric scalar potential with a bidoublet and left and right doublets, for given $k$, are
\begin{equation}
\begin{split}
  \lambda_{L} > 0,
  \quad
  \lambda_{R} > 0,
  \quad
  \lambda_{00} &> 0, 
  \\  
  \bar{\lambda}_{LR} = \frac{1}{2} \lambda_{LR} + \sqrt{\lambda_{L} \lambda_{R}} > 0,
  \quad
  \bar{\lambda}_{L0} = \frac{1}{2} \lambda_{L0} + \sqrt{\lambda_{L} \lambda_{00}} > 0,
  \quad
  \bar{\lambda}_{R0} = \frac{1}{2} \lambda_{R0} + \sqrt{\lambda_{R} \lambda_{00}} &> 0,
  \\
  \sqrt{\lambda_{L} \lambda_{R} \lambda_{00}} + \lambda_{LR} \sqrt{\lambda_{00}} + \lambda_{L0} \sqrt{\lambda_{R}} + \lambda_{R0} \sqrt{\lambda_{L}} 
  + \sqrt{2 \bar{\lambda}_{LR} \bar{\lambda}_{L0} \bar{\lambda}_{R0}} &> 0.
\end{split}
\label{eq:coupling:real:vacuum:left:right:ness}
\end{equation}

Alternatively, we write down the quartic couplings in the basis $(r^{0}, r^{1}, \abs{H_{L}}^{2}, \abs{H_{R}}^{2})$ and rotate the $1+1$ forward lightcone $\text{LC}^{+}$ into the non-negative quadrant $\mathbb{R}^{2}_{+}$ of the $r^{0}r^{1}$-plane. The resulting orbit space is the non-negative orthant $\mathbb{R}_{+}^{4}$ and therefore we can apply copositivity \cite{Kannike:2012pe} to the obtained quartic coupling matrix, given by
\begin{equation}
  \lambda = 
  \begin{pmatrix}
    \lambda_{-} & \frac{1}{2} \lambda_{\mp} & \frac{1}{2} \lambda_{L-} & \frac{1}{2} \lambda_{R-}
    \\
    \frac{1}{2} \lambda_{\mp} & \lambda_{+} & \frac{1}{2} \lambda_{L+} & \frac{1}{2} \lambda_{R+}
    \\
    \frac{1}{2} \lambda_{L-} & \frac{1}{2} \lambda_{L+} & \lambda_{L} & \frac{1}{2} \lambda_{LR}
    \\
    \frac{1}{2} \lambda_{R-} & \frac{1}{2} \lambda_{R+} & \frac{1}{2} \lambda_{LR} & \lambda_{R}
  \end{pmatrix},
\end{equation}
where
\begin{align}
  \lambda_{-} &= \frac{1}{2} (\lambda_{1} + 2 \lambda_{2} + \lambda_{3} - 2 \lambda_{4}),
  \\
  \lambda_{+} &= \frac{1}{2} (\lambda_{1} + 2 \lambda_{2} + \lambda_{3} + 2 \lambda_{4}),
  \\
  \lambda_{\mp} &= \lambda_{1} - 2 \lambda_{2} - \lambda_{3} + 2 (\lambda_{3} - 2 \lambda_{2}) \theta(2 \lambda_{2} - \lambda_{3}),
  \label{eq:lambda:mp}
  \\
  \lambda_{L-} &= \frac{1}{\sqrt{2}} [\lambda_{\Phi L} + \rho_{\Phi L} \lambda'_{\Phi  L} + (1 -\rho_{\Phi L}) \tilde{\lambda}'_{\Phi  L} - 2 \tilde{\lambda}_{\Phi L}],
  \\
  \lambda_{R-} &=  \frac{1}{\sqrt{2}} [\lambda_{\Phi R} + \rho_{\Phi R} \lambda'_{\Phi  R} + (1 -\rho_{\Phi R}) \tilde{\lambda}'_{\Phi  R} - 2 \tilde{\lambda}_{\Phi R}],
  \\
 \lambda_{L+} &= \frac{1}{\sqrt{2}} [\lambda_{\Phi L} + \rho_{\Phi L} \lambda'_{\Phi  L} + (1 -\rho_{\Phi L}) \tilde{\lambda}'_{\Phi  L} + 2 \tilde{\lambda}_{\Phi L}],
  \\
  \lambda_{R+} &=  \frac{1}{\sqrt{2}} [\lambda_{\Phi R} + \rho_{\Phi R} \lambda'_{\Phi  R} + (1 -\rho_{\Phi R}) \tilde{\lambda}'_{\Phi  R} + 2 \tilde{\lambda}_{\Phi R}],
\end{align}
where we have taken into account the relations \eqref{eq:rho:rel} to substitute for $\tilde{\rho}_{\Phi L}$ and $\tilde{\rho}_{\Phi R}$. We must consider all combinations of values $(\rho_{\Phi L}, \rho_{\Phi R}) = (0,0), (1,0), (0,1), (1,1)$.
The resulting sufficient vacuum stability conditions for the left-right symmetric scalar potential with a bidoublet and left and right doublets are given by
\begin{equation}
\begin{split}
  \lambda_{L} > 0,
  \quad
  \lambda_{R} > 0,
  \quad
  \lambda_{-} > 0, 
  \quad
  \lambda_{+} &> 0,
  \\
  \bar{\lambda}_{LR} = \frac{1}{2} \lambda_{LR} + \sqrt{ \lambda_{L} \lambda_{R} } > 0,
  \quad
  \bar{\lambda}_{L-} = \frac{1}{2} \lambda_{L-} + \sqrt{ \lambda_{L} \lambda_{-} } > 0,
  \quad
  \bar{\lambda}_{L+} = \frac{1}{2} \lambda_{L+} +\sqrt{ \lambda_{L} \lambda_{+} } &> 0,
  \\
  \bar{\lambda}_{R-} = \frac{1}{2} \lambda_{R-} + \sqrt{ \lambda_{R} \lambda_{-} } > 0,
  \quad
  \bar{\lambda}_{R+} = \frac{1}{2} \lambda_{R+} +\sqrt{ \lambda_{R} \lambda_{+} } > 0,
  \quad
  \bar{\lambda}_{\mp} = \frac{1}{2} \lambda_{\mp} + \sqrt{ \lambda_{-} \lambda_{+} } &> 0,
   \\
  \sqrt{\lambda_{L} \lambda_{R} \lambda_{-}} + \lambda_{LR} \sqrt{\lambda_{-}} + \lambda_{L-} \sqrt{\lambda_{R}} + \lambda_{R-} \sqrt{\lambda_{L}} 
  + \sqrt{2 \bar{\lambda}_{LR} \bar{\lambda}_{L-} \bar{\lambda}_{R-}} &> 0,
     \\
  \sqrt{\lambda_{L} \lambda_{R} \lambda_{+}} + \lambda_{LR} \sqrt{\lambda_{+}} + \lambda_{L+} \sqrt{\lambda_{R}} + \lambda_{R+} \sqrt{\lambda_{L}} 
  + \sqrt{2 \bar{\lambda}_{LR} \bar{\lambda}_{L+} \bar{\lambda}_{R+}} &> 0,
  \\
  \sqrt{\lambda_{L} \lambda_{-} \lambda_{+}} + \lambda_{L-} \sqrt{\lambda_{+}} + \lambda_{L+} \sqrt{\lambda_{-}} + \lambda_{\mp} \sqrt{\lambda_{L}} 
  + \sqrt{2 \bar{\lambda}_{L-} \bar{\lambda}_{L+} \bar{\lambda}_{\mp}} &> 0,
  \\
  \sqrt{\lambda_{R} \lambda_{-} \lambda_{+}} + \lambda_{R-} \sqrt{\lambda_{+}} + \lambda_{R+} \sqrt{\lambda_{-}} + \lambda_{\mp} \sqrt{\lambda_{R}} 
  + \sqrt{2 \bar{\lambda}_{R-} \bar{\lambda}_{R+} \bar{\lambda}_{\mp}} &> 0,
  \\
  \det (\lambda) > 0 \lor \text{some element(s) of} \adj (\lambda) &< 0,
\end{split}
\label{eq:coupling:real:vacuum:left:right:suff}
\end{equation}
where the last condition, obtained from the Cottle-Habetler-Lemke theorem \cite{Cottle1970295}, is not given in full. The adjugate $\adj(A)$ of a matrix $A$ is the transpose of the cofactor matrix of $A$. It is defined through the relation $A \adj(A) = \det(A) \, I$.

\section{Conclusions}
\label{sec:concl}

Finding vacuum stability conditions and minimising scalar potentials is an involved problem. It helps to consider the potential as a function of gauge invariants, not directly of scalar field components. The orbit space of these invariants has a peculiar shape dependent on the group representations. In particular, the 2HDM orbit space resembles a forward lightcone in $1+3$ dimensions \cite{Ivanov:2006yq,Maniatis:2006fs}.

We show that the orbit space of a scalar field in a complex square matrix representation, with two quadratic invariants, also has a lightcone shape in $1+2$ dimensions. The Minkowski space structure is a parametrisation of the Cauchy-Schwarz inequality \eqref{eq:matrix:Cauchy-Schwarz} for the matrix inner product. Positivity of the quartic coupling tensor of the matrix field gives the vacuum stability conditions \eqref{eq:pos:cond:gen:1+1:1}, \eqref{eq:pos:cond:gen:1+1:2} and \eqref{eq:pos:cond:gen:1+1:3} for its self-couplings. The method is suitable for treating scalar fields such as a bidoublet, a complex $SU(2)$ triplet or a complex $SU(3)$ octet. For a realistic model, portal couplings to the Higgs doublet need to be included. In this case, finding the vacuum stability conditions becomes complicated, but in the most interesting case of real couplings, it can be reduced to copositivity, yielding the simple conditions \eqref{eq:gen:cop}. Minima of the potential can be found in the same formalism.

Finally, we use the formalism to derive the necessary vacuum stability conditions \eqref{eq:coupling:real:vacuum:left:right:ness} and the sufficient vacuum stability conditions \eqref{eq:coupling:real:vacuum:left:right:suff} for the quartic couplings of a left-right symmetric potential of a bidoublet and left and right Higgs doublets.

\appendix

\section{Derivatives of the lightcone variables}
\label{sec:deriv}

The derivatives of the lightcone variables with respect to the scalar field $M$ are given by
\begin{equation}
  \frac{\partial r^{0}}{\partial M} = M^{*}, \quad \frac{\partial r^{1}}{\partial M} = M^{T}, \quad \frac{\partial r^{2}}{\partial M} = \frac{1}{i} M^{T},
  \label{eq:d:r:d:M}
\end{equation}
which can be taken into invariant form by multiplying by either $M^{T}$ or $M^{*}$ and 
taking a trace:

\begin{align}
  \tr  \frac{\partial r^{0}}{\partial M} M^{T} &= r^{0}, 
  &
  \tr  \frac{\partial r^{0}}{\partial M} M^{*} &= r^{1} - i r^{2}, 
  \label{eq:d:r:d:M:M:1}
  \\
  \tr  \frac{\partial r^{1}}{\partial M} M^{T} &= r^{1} + i r^{2},
  & 
  \tr  \frac{\partial r^{1}}{\partial M} M^{*} &= r^{0},
  \label{eq:d:r:d:M:M:2}
  \\
  \tr  \frac{\partial r^{2}}{\partial M} M^{T} &= -i r^{1} + r^{2},
  & \tr  \frac{\partial r^{2}}{\partial M} M^{*} &= -i r^{0}.
  \label{eq:d:r:d:M:M:3}
\end{align}

\section{Minimum Solutions}
\label{sec:minima}

For completeness, we present the extremum solutions discussed in Section~\ref{sec:minim} in detail. We assume $\lambda'_{M} - \lambda''_{M} \geqslant 0$ and take $r^{2} = 0$. If $\lambda'_{M} - \lambda''_{M} < 0$, one has to take $(r^{2})^{2} = (r^{0})^{2} - (r^{1})^{2}$ and let $\lambda_{M} \to \lambda_{M} + \lambda'_{M} - \lambda''_{M}$ and $\lambda'_{M} + \lambda''_{M} \to \lambda'_{M} + \lambda''_{M} - (\lambda'_{M} - \lambda''_{M}) = 2 \lambda''_{M}$. 

The trivial solution is given by
\begin{equation}
  \abs{H}^{2} = r^{0} = r^{1} = 0.
\end{equation}
If only the Higgs doublet has a vacuum expectation value (VEV), then
\begin{equation}
  \abs{H}^{2} = -\frac{\mu^{2}_{H}}{2 \lambda_{H}}, \quad r^{0} = r^{1} = 0.
\end{equation}
The solutions to Eq.~\eqref{eq:pm:sols} are given by

\begin{equation}
  \abs{H}^{2} = 0, \quad r^{0} = - \frac{1}{2} \frac{\mu^{2}_{M} \pm \mu^{\prime 2}_{M}}{\lambda_{M} + \lambda'_{M} + \lambda''_{M} \pm \lambda'''_{M}}, \quad r^{1} = \pm r^{0}
\end{equation}
and
\begin{equation}
\begin{split}
  \abs{H}^{2} &= -\frac{2 (\lambda_{M} + \lambda'_{M} + \lambda''_{M} \pm \lambda'''_{M})
\mu^{2}_{H} - (\lambda_{HM} \pm \lambda'_{HM}) (\mu^{2}_{M} \pm \mu^{\prime 2}_{M})}{4 \lambda_{H} (\lambda_{M} + \lambda'_{M} + \lambda''_{M} \pm \lambda'''_{M})
 - (\lambda_{HM} \pm \lambda'_{HM})^2}, 
\\
r^{0} &= -\frac{
   2 \lambda_{H} (\mu^{2}_{M} \pm \mu^{\prime 2}_{M}) - (\lambda_{HM} \pm \lambda'_{HM}) 
\mu^{2}_{H}}{4 \lambda_{H} (\lambda_{M} + \lambda'_{M} + \lambda''_{M} \pm \lambda'''_{M})
 - (\lambda_{HM} \pm \lambda'_{HM})^2},
\\
r^{1} &= \pm r^{0}.
\end{split}
\end{equation}
The solutions to Eq.~\eqref{eq:r:mu:eq} are given by
\begin{equation}
  \abs{H}^{2} = 0, \quad r^{0} = \frac{\lambda'''_{M} \mu^{\prime 2}_{M} - 2 (\lambda'_{M} + \lambda''_{M}) \mu^{2}_{M}}{4 \lambda_{M} (\lambda'_{M} + \lambda''_{M}) - \lambda_{M}^{\prime\prime\prime 2}}, 
  \quad  
  r^{1} = \frac{\lambda'''_{M} \mu^{2}_{M} - 2 \lambda_{M} \mu^{\prime 2}_{M}}{4 \lambda_{M} (\lambda'_{M} + \lambda''_{M}) - \lambda_{M}^{\prime\prime\prime 2}}
\end{equation}
and
\begin{equation}
\begin{split}
  \abs{H}^{2} &= \frac{1}{2} [   
    4 \lambda_{M} (\lambda'_{M} + \lambda''_{M}) \mu^{2}_{H}
 - 
\lambda_{M}^{\prime\prime\prime 2} \mu^{2}_{H} -2 \lambda_{M} \lambda'_{HM} \mu^{\prime 2}_{M} - 
    2 (\lambda'_{M} + \lambda''_{M}) \lambda_{HM} \mu^{2}_{M} 
    \\
    &+
    \lambda'''_{M} (\lambda_{HM} \mu^{\prime 2}_{M} + \lambda'_{HM} \mu^{2}_{M})]
    /[
   (\lambda'_{M} + \lambda''_{M}) \lambda_{HM}^2 - 
\lambda'''_{M} \lambda_{HM} \lambda'_{HM} + \lambda_{M} \lambda_{HM}^{\prime 2} 
  \\
  &+ (\lambda_{M}^{\prime\prime\prime 2} -4 \lambda_{M} (\lambda'_{M} + \lambda''_{M})) 
\lambda_{H}],
\\
  r^{0} &= \frac{1}{2} [4 \lambda_{H} (\lambda'_{M} + \lambda''_{M})  \mu^{2}_{M} - \lambda_{HM}^{\prime 2} \mu^{2}_{M} + \lambda_{HM} \lambda'_{HM} \mu^{\prime 2}_{M}  - 
    2 (\lambda'_{M} + \lambda''_{M}) \lambda_{HM} \mu^{2}_{H} 
  \\
  &+ 
\lambda'''_{M} (\lambda'_{HM} \mu^{2}_{H} -2 \lambda_{H} \mu^{\prime 2}_{M})]/[
   (\lambda'_{M} + \lambda''_{M}) \lambda_{HM}^2 
   -\lambda'''_{M} \lambda_{HM} \lambda'_{HM} + \lambda_{M} \lambda_{HM}^{\prime 2} 
  \\
  &+ 
(\lambda_{M}^{\prime\prime\prime 2} -4 \lambda_{M} (\lambda'_{M} + \lambda''_{M})) 
\lambda_{H}],
  \\
  r^{1} &= \frac{1}{2} [-\lambda_{HM}^2 \mu^{\prime 2}_{M} + 
    4 \lambda_{M} \lambda_{H} \mu^{\prime 2}_{M} + \lambda_{HM} \lambda'_{HM} 
\mu^{2}_{M} - 
    \lambda'''_{M} (2  \lambda_{H} \mu^{2}_{M} + \lambda_{HM} 
\mu^{2}_{H}) - 2 \lambda_{M} \lambda'_{HM} \mu^{2}_{H}]
\\
&/[
   (\lambda'_{M} + \lambda''_{M}) \lambda_{HM}^2 - 
\lambda'''_{M} \lambda_{HM} \lambda'_{HM} + \lambda_{M} \lambda_{HM}^{\prime 2} + 
(\lambda_{M}^{\prime\prime\prime 2} -4 \lambda_{M} (\lambda'_{M} + \lambda''_{M})) 
\lambda_{H}].
\end{split}
\end{equation}

\section*{Acknowledgments}
We woud like to thank Luca Marzola for useful discussions and for reading the draft of the manuscript. This work was supported by the Estonian Research Council grant PRG434, by the European Regional Development Fund and the programme Mobilitas Pluss grant MOBTT5, and by the EU through the European Regional Development Fund CoE program TK133 ``The Dark Side of the Universe". 

\bibliographystyle{JHEP}
\bibliography{lightcone}

\end{document}